\documentclass[sigconf,nonacm,screen]{acmart}
\usepackage{url}
\usepackage{xspace}
\usepackage{multirow}

\newcommand{\etal}{\textit{et al.}\xspace}
\newcommand{\ie}{\textit{i.e.}\xspace}

\newcommand{\Heading}[1]{\textbf{#1.}}
\newcommand{\RQ}[1]{\textit{RQ}${}_{\mathrm{#1}}$}
\usepackage{framed}
\newcommand{\Conclusion}[1]{\begin{framed}\noindent #1\end{framed}}
\newcommand{\W}[1]{\textbf{#1}}

\def\support{\mathit{support}}
\def\confidence{\mathit{confidence}}

\def\minsup{\textit{minsup}\xspace}
\def\minconf{\textit{minconf}\xspace}
\def\MapAll{$\mathsf{MAP}_\mathit{all}$\xspace}
\def\MapApp{$\mathsf{MAP}_\mathit{app}$\xspace}
\def\Precision{\textit{Precision}\xspace}

\def\Full{\textit{Full}\xspace}
\def\FirstParent{\textit{First-Parent}\xspace}
\def\FirstParentM{$\text{\textit{First-Parent}}_\mathrm{merge}$\xspace}
\def\FirstParentNM{$\text{\textit{First-Parent}}_\mathrm{no\_merge}$\xspace}
\def\ImpNM{$\mathit{Imp}_\mathrm{no\_merge}$\xspace}
\def\ImpM{$\mathit{Imp}_\mathrm{merge}$\xspace}

\def\ApacheK{$\mathit{Apache}_K$\xspace}
\def\EclipseK{$\mathit{Eclipse}_K$\xspace}
\def\ApacheM{$\mathit{Apache}_M$\xspace}

\title[Revisiting the Effect of Branch Handling Strategies on Change Recommendation]%
{Revisiting the Effect of Branch Handling Strategies\\on Change Recommendation}

\author{Keisuke Isemoto}
\email{k\_isemoto@se.c.titech.ac.jp}
\affiliation{%
  \institution{Tokyo Institute of Technology}
  \streetaddress{Ookayama 2--12--1}
  \city{Meguro-ku}
  \state{Tokyo}
  \country{Japan}
  \postcode{152--8550}
}
\author{Takashi Kobayashi}
\email{tkobaya@c.titech.ac.jp}
\affiliation{%
  \institution{Tokyo Institute of Technology}
  \streetaddress{Ookayama 2--12--1}
  \city{Meguro-ku}
  \state{Tokyo}
  \country{Japan}
  \postcode{152--8550}
}
\author{Shinpei Hayashi}
\email{hayashi@c.titech.ac.jp}
\affiliation{%
  \institution{Tokyo Institute of Technology}
  \streetaddress{Ookayama 2--12--1}
  \city{Meguro-ku}
  \state{Tokyo}
  \country{Japan}
  \postcode{152--8550}
}

\begin{abstract}
  Although literature has noted the effects of branch handling strategies on change recommendation based on evolutionary coupling, they have been tested in a limited experimental setting.
  Additionally, the branches characteristics that lead to these effects have not been investigated.
  In this study, we revisited the investigation conducted by Kovalenko \etal on the effect to change recommendation using two different branch handling strategies: including changesets from commits on a branch and excluding them.
  In addition to the setting by Kovalenko \etal, we introduced another setting to compare: extracting a changeset for a branch from a merge commit at once.
  We compared the change recommendation results and the similarity of the extracted co-changes to those in the future obtained using two strategies through 30 open-source software systems.
  The results show that handling commits on a branch separately is often more appropriate in change recommendation, although the comparison in an additional setting resulted in a balanced performance among the branch handling strategies.
  Additionally, we found that the merge commit size and the branch length positively influence the change recommendation results.
\end{abstract}

\ccsdesc[500]{Software and its engineering}
\ccsdesc[500]{Software and its engineering~Software version control}
\ccsdesc[500]{Software and its engineering~Software evolution}

\keywords{Change recommendation, evolutionary coupling, version control, Git}

\begin{document}
  
\maketitle

\section{Introduction}\label{s:introduction}

As software systems evolve, the amount of codes and the dependencies between the codes increase, thus making the systems more complex.
As their complexity increases, it is difficult for developers to understand the extent and effect of their changes.
Therefore, necessary changes can leak out, causing bugs in the source code.

Dependencies among source files can be identified based on information about the evolutionary processes of the source files.
By using information such as commits recorded in version control systems, we can estimate dependencies that cannot be obtained through static or dynamic analysis, independent of the used programming language.

One approach is change recommendation based on evolutionary coupling~\cite{rolfsnes2016generalizing,zimmermann2005mining}.
The co-change information can be obtained from the revision history of the source files that have been changed in the same revision.
By applying association rule mining to the co-change information, the dependencies among source files can be extracted as association rules such as ``\textit{when a file $f_1$ is changed, another file $f_2$ should also be changed}.''

Researchers have noted that there are various factors that affect the results of change recommendations based on co-change information~\cite{moonen2016practical,moonen2018effects}.
Kovalenko \etal reported the influence of branches in history~\cite{kovalenko2018mining}.
They compared two history extraction approaches with different branch handling strategies, and reported that obtaining co-change information from commits on a branch slightly improved the recommendation performance compared to omitting to obtain it from history.

However, unlike the result obtained by Kovalenko \etal~\cite{kovalenko2018mining}, another possible approach when extracting co-change information from history is to obtain such co-change information collectively from a merge commit without traversing commits on a branch.
A similar study conducted by Miura \etal~\cite{miura2016impact} concluded that changes should be summarized by \emph{work item} in co-change analysis, and it is unclear how the changes on a branch should be handled when applying a co-change analysis.
Miura \etal investigated the co-change information extracted from the version history of open-source software systems, and found that the changes in commits identified as following the same work, \ie, those related to the same issue, should be grouped together to avoid missing co-change information~\cite{miura2016impact}.
Branches in version control systems are primarily used for achieving specific tasks, such as bug fixes and feature additions~\cite{zou2019branch}, and changes on a branch are comparable to the same work.
Based on this observation, an approach skipping commits on a branch but obtaining their co-change information from a merge commit is possible, and the appropriate treatment of branches when extracting changes during change recommendation is unclear.

Additionally, because branches are an essential element of software development, it is important to investigate the influence of branches in applying repository mining methods to actual repositories.
Currently, Git is widely used as a representative version control system.
In projects that use Git, change management using branches is widely practiced~\cite{bird2012assessing}.
The branches in Git are easy to create, can be used for tasks such as fixing bugs or adding features~\cite{zou2019branch}, and can be used to manage each release.
Branch usage also affects the granularity of commits and change recommendations; hence, branches need to be handled appropriately depending on the characteristics of the target repository.
Additionally, investigating the types of branches that have affected the change recommendation should be helpful in implementing repository mining methods in actual development.

In this study, we set the following research questions~(RQs) to identify the appropriate method for handling branches in change recommendations.
\def\RQone{Does the experimental setting of the study on branch handling strategies influence the change recommendation performance?}
\def\RQtwo{Which branches characteristics affect the performance of change recommendations?}
\def\RQtwoone{What is the relationship between the branch characteristics and change recommendation results?}
\def\RQtwotwo{What is the relationship between the branch characteristics and the co-changes extracted from commits in a branch and those from a merge commit?}
\begin{itemize}
  \item \RQ{1}: \RQone
  \item \RQ{2}: \RQtwo
  \begin{itemize}
    \item \RQ{2.1}: \RQtwoone
    \item \RQ{2.2}: \RQtwotwo
  \end{itemize}
\end{itemize} 

To answer \RQ{1}, we compare the change recommendation results with the two branch handling strategies proposed by Kovalenko \etal~\cite{kovalenko2018mining} to investigate the effect of the experimental setting that might lead to the results.
We used more repositories than Kovalenko \etal as well as two implementations.
The first implementation reproduced Kovalenko \etal's study, and the other implementation changed the handling of merge commits.
We compared the results with different branch handling strategies, and analyzed them through validation based on more evaluation metrics than Kovalenko \etal

In \RQ{2}, we investigate the branches characteristics that affect the change recommendations.
However, because differences exist among the branches, defining a uniform treatment for the branches might not be possible.
Therefore, we investigate how each branch characteristic affects the change recommendation, such as the length of branches and the size of the merge commits.
In \RQ{2.1}, we analyze the differences in the recommendation results of different branch handling strategies for each branch characteristic.
Because the analysis in answering \RQ{2.1} can be affected by the configuration of the change recommendation method used, in \RQ{2.2} we also analyze the difference in co-change information directly between the commits on a branch and those in a merge commit for each branch characteristic.

The rest of this paper is organized as follows.
We first briefly explain the change recommendation approach based on evolutionary coupling in the next section.
In Section~\ref{s:prior-study}, we explain the study by Kovalenko \etal and introduce RQs.
We respond to the introduced RQs in Section~\ref{s:empirical}.
Discussion and implications, and threats to validity are described in Sections~\ref{s:discussion} and \ref{s:threats}, respecctively.
Sections~\ref{s:related} and \ref{s:conclusion} are for related work and concluding remarks.

\section{Change Recommendation based on Evolutionary Coupling}\label{s:preliminary}

Change recommendation approaches based on evolutionary coupling~\cite{rolfsnes2016generalizing,zimmermann2005mining,kovalenko2018mining} take a set of files as an input query, and output a ranked list of files as recommendations to be fixed in addition to the input set of files.
In this study, we use the ``other files'' algorithm, which is a change recommendation algorithm implemented by Kovalenko \etal~\cite{kovalenko2018appendix}.
In the following, we briefly explain this as a typical method of co-change-based change recommendation.

This change recommendation algorithm comprises the following three steps:
\begin{enumerate}
  \item Collection of commits to be used.
  \item Application of the \textit{Apriori} algorithm~\cite{apriori} to the changesets extracted from the collected commits to generate \emph{association rules}.
  \item Construction of a \emph{recommendation list} of files using the obtained association rules.
\end{enumerate}

The first step is to collect commits from the target change repository using the given branch handling strategy.
Each commit is regarded as a \emph{changeset}, \ie, a set of files that were modified at the commit.
Initially, commits that meet the following inclusion criteria are collected by following the given branch handling strategy.
\begin{itemize}
  \item Only commits that contain at least one file in the query are included.
  \item Assuming that changed files in large commits may not have meaningful relationships~\cite{moonen2016practical}, commits with more than 10 changed files are excluded.
\end{itemize}
Finally, at most, 100 latest files are passed onto the next step.

The second step is to generate \emph{association rules}.
The set of changed files in each collected commit forms a \emph{transaction}; \ie, each pair of files in the changeset is changed together.
The transaction database $T$, which is a set of transactions obtained from the commits, is the input for the Apriori algorithm for generating association rules as the recommendation results.
Each association rule follows the form of ``\textit{condition part} $\to$ \textit{conclusion part}'' ($x \to y$), where each part is a set of files.
The quality of an association rule is measured by the \emph{support} and \emph{confidence} metrics:
\begin{align*}
  \support(x)          &= |\{\, t \in T \mid x \subseteq t \,\}| / |T|, \\
  \support(x \to y)    &= \support(x \cup y), \\
  \confidence(x \to y) &= \support(x \to y) / \support(x).
\end{align*}
In the Apriori algorithm, two parameters, the minimum support~(\minsup) and confidence~(\minconf), are employed to generate the association rules that meet the conditions from the given transaction database.
We used the same parameters as those in Kovalenko \etal: (\minsup, \minconf) = (0.1, 0.1).
The generated association rules having more than one file at the conclusion part are filtered out.
Finally, at most, 10 association rules having top support values are passed onto the next step.

The third step is to produce a \emph{recommendation list} by applying the given query to the obtained association rules.
If all the files in the condition part of an association rule are included in the query, the file at the conclusion part of the rule is adopted as a recommendation result.
The results in the recommendation list are ranked in the order of support of the adopted rules.

In this change recommendation approach, the number of recommendations is kept low, because the association rules to be used are narrowed down before making the recommendation.
Because the recommendation is made from at most 10 association rules, the resulting recommended files are also at most 10.

\section{Branch Handling Strategies}\label{s:prior-study}

\begin{figure}[t]\centering
  \includegraphics[width=7.5cm]{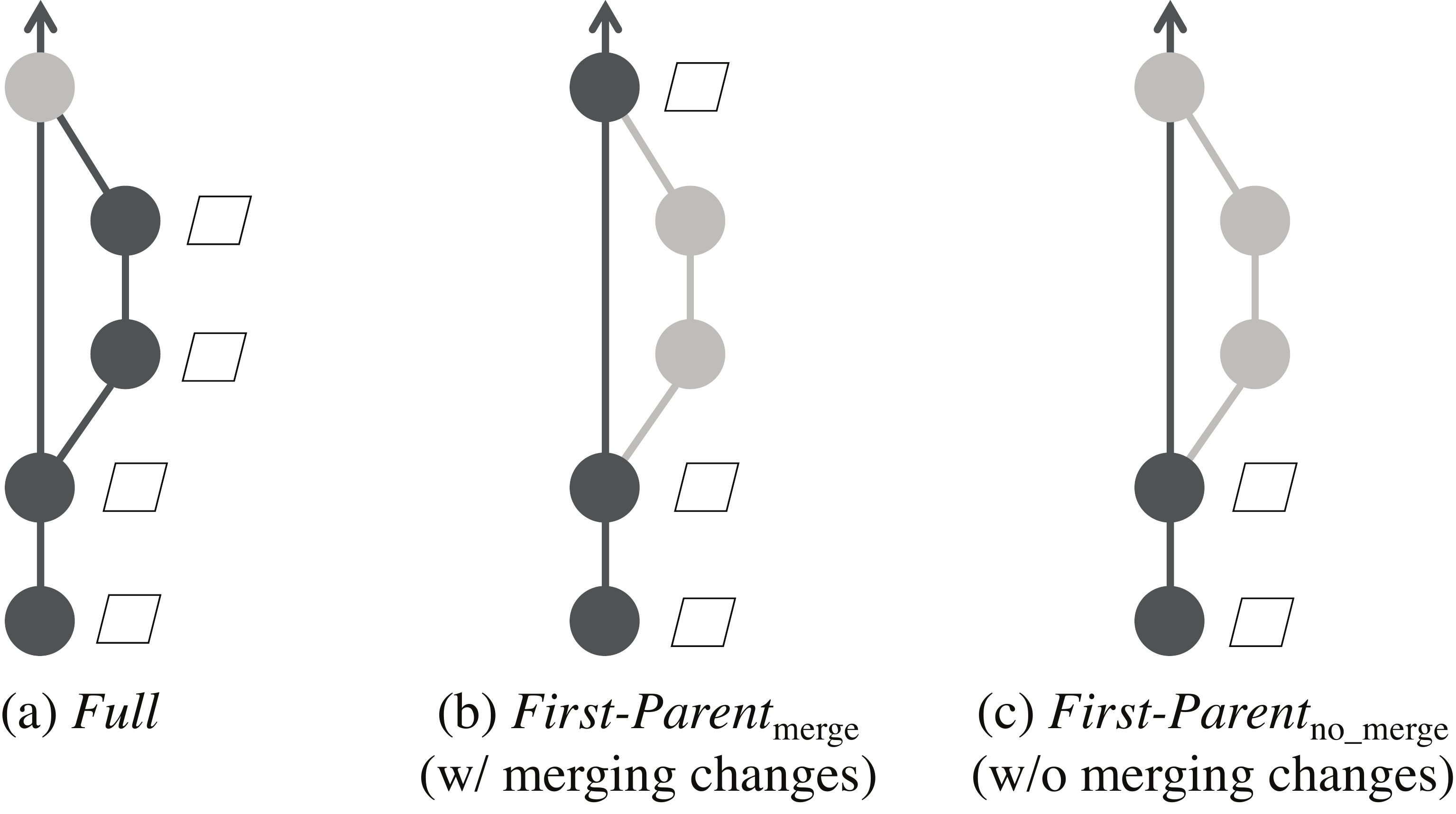}
  \caption{Branch handling strategies.}\label{f:historyFlow}
\end{figure}

Kovalenko \etal~\cite{kovalenko2018mining} studied the effect of handling branches when applying repository mining tasks to a Git repository.
They investigated three repository mining applications: recommending reviewers~\cite{Balachandran2013Reviewer}, bug prediction~\cite{hassan2009predicting}, and change recommendation~\cite{zimmermann2005mining}.
They compared the results after applying two branch handling strategies, \FirstParent and \Full, to three repository mining methods, including change recommendation, to evaluate the importance of each branch handling strategy.
Figure~\ref{f:historyFlow} shows each branch handling strategy.
The commits with a parallelogram are the targets to be extracted in the branch handling strategy used.
The \FirstParent strategy selects commits by tracing the first parent when a merge commit occurs and omits the changes obtained from non-main branches, whereas the \Full strategy selects commits that are reachable by tracing all parent commits.
The \Full strategy excludes the merging change in merge commits, as they are a union of all changes on the associated branch.
The \FirstParent strategy defined by Kovalenko \etal also excludes the merging change in merge commits to compare the effect of using the changes in branches.
For clarity, we named this strategy as \FirstParentNM.
Kovalenko \etal compared these two branch handling strategies, and demonstrated that the \Full strategy exhibited a slightly better recommendation performance.

Nevertheless, there is another possible approach for handling changes in a branch: extracting all the changes from a merge commit to obtain co-change information of a branch collectively.
We name this strategy \FirstParentM.
In the \Full strategy, each commit on a branch is treated individually.
In the \FirstParentM strategy, the changes on the branch are combined into a merge commit, and a single changeset from the merge commit is used.
In contrast to the \FirstParentM strategy, the \FirstParentNM strategy excludes merge commits with some exceptions; only the cases where the merge commits contained additional changes are included.
The difference of handling merge commits might affect the change recommendation performance, wherein the performance of the \Full strategy tended to be higher.

Therefore, in \RQ{1}, we compare the performance of the change recommendation results between \Full and \FirstParentNM and between \Full and \FirstParentM.
Through experiments with the two combinations, we confirmed the influence of missing changes on the branches as Kovalenko \etal studied.
By experimenting with \Full and \FirstParentNM, we can experiment with almost the same conditions as Kovalenko \etal, leading to a fairer comparison.
Additionally, \RQ{2} considers the missing changes in the branches of the \FirstParent strategy to be unsuitable for comparing the branch handling; hence, we target the recommendation results between the \Full and \FirstParentM strategies.

\begin{figure*}[t]\centering
  \includegraphics[width=11cm]{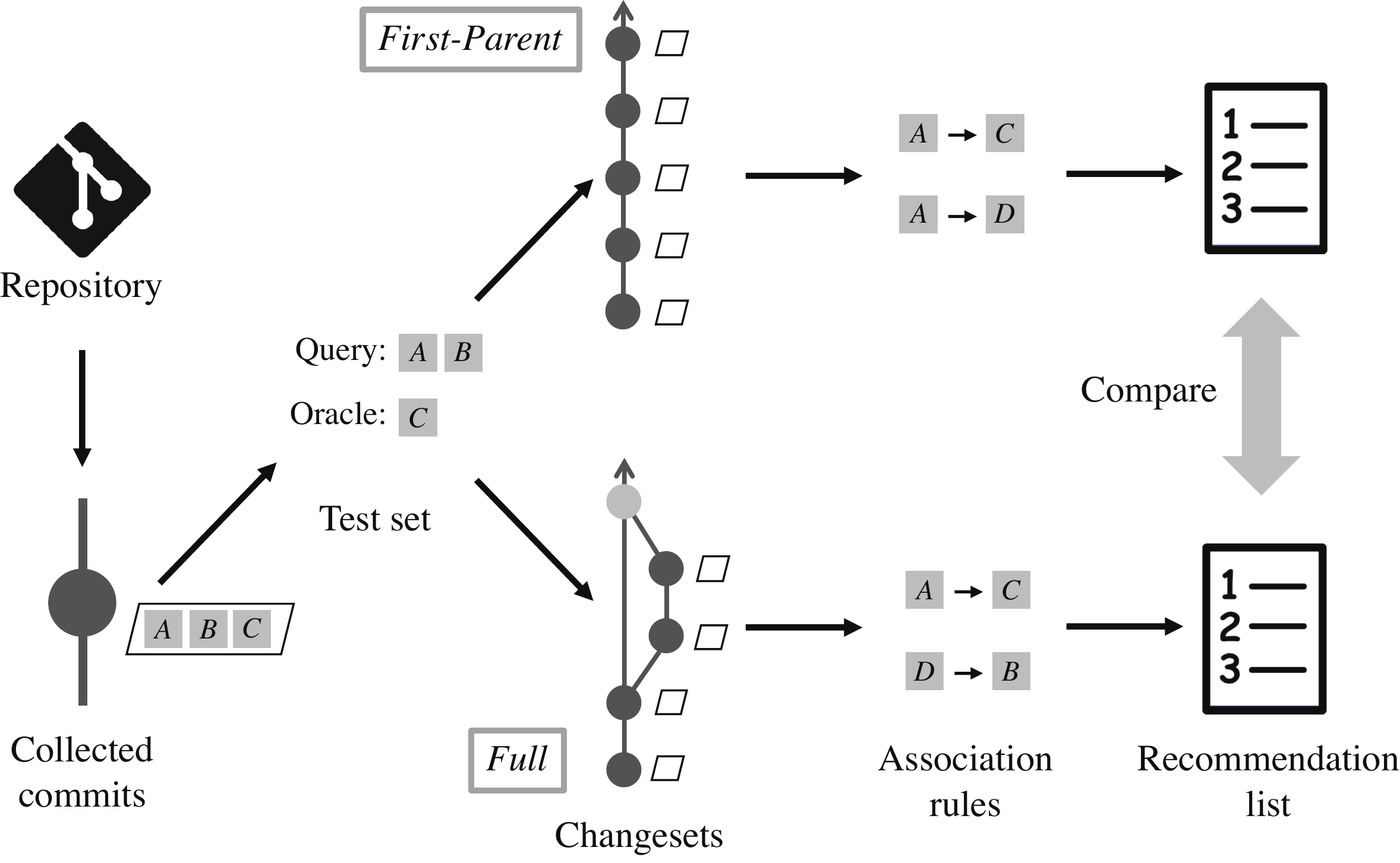}
  \caption{Comparison of two change recommendation results obtained using different branch handling strategies.}\label{f:expFlow}
\end{figure*}

\section{Empirical Studies}\label{s:empirical}

\begin{table}[t]\centering
  \caption{Target repository sets}\label{t:repoList}
  \begin{tabular}{c|rr} \hline
    Repository sets & \# repositories & Avg.\ \# commits \\ \hline
    \ApacheK        &              10 &          3,955.6 \\
    \EclipseK       &               8 &         15,582.9 \\
    \ApacheM        &              12 &          2,829.6 \\ \hline
  \end{tabular}
\end{table}

\subsection{Dataset Preparation}

In this study, we used the repository sets of the Apache Software Foundation~(ASF) and Eclipse ecosystems used by Kovalenko \etal~\cite{kovalenko2018mining}~(\ApacheK and \EclipseK, respectively) and that of the ASF ecosystem used by Miura \etal~\cite{miura2016impact}~(\ApacheM), comprising 30 repositories in total.
Table~\ref{t:repoList} presents the statistics of the used repository sets.
Note that two repositories from the Eclipse ecosystem, which failed to clone, and two repositories from ASF, which were unaffected by the branch handling strategy selection, were excluded.

\subsection{\RQ{1}: \RQone}

\subsubsection{Motivation}

Kovalenko \etal~\cite{kovalenko2018mining} compared the change recommendation results using two branch handling strategies, and they reported that the \Full strategy slightly outperformed the \FirstParentNM strategy.
In contrast, according to Miura \etal~\cite{miura2016impact}, the co-change information should be handled at the task~(work item) level in the evolutionary coupling analysis because the revision~(commit) level analysis fails to extract the co-change information over revisions within the task.
If we assume that the changes made on a single branch are all related to a single task, there may lead to another strategy to extract all the changes in a branch from a merge commit at once: \FirstParentM.
Miura \etal~\cite{miura2016impact} found that missing co-changes could be prevented by employing the task level change treatment, while Kovalenko \etal found that the \Full strategy, which treats commits on a branch separately, produced both a higher recommendation success rate and higher number of recommendations.
Because the branches are mainly used as implementing features, \ie, as feature branches~\cite{zou2019branch}, if we consider that the changes on a branch are related to a task, summarizing them at the task level is similar to merging the changes on a branch.
Although the study by Kovalenko \etal compared the performance of the \Full and \FirstParentNM strategies, a different comparison with \FirstParentM can lead to understanding branch handling strategies more on their change recommendation performance.
The \FirstParentNM strategy skips many changes on the branches, which may have resulted in the slightly higher performance of the \Full strategy.
If the performance of the \Full strategy was higher because of the co-change information being excluded, it does not necessarily mean that the performance of the \Full strategy is better than that of the \FirstParentM strategy in general.
Therefore, we experimented with an implementation that follows Kovalenko \etal~\cite{kovalenko2018mining}, denoted as the \FirstParentNM strategy.
We also used another implementation to process all the changes in the collected merge commits, denoted as the \FirstParentM strategy.

Additionally, Kovalenko \etal used the success rate and number of recommendations for the evaluation metrics, which might be room for improvement.
For example, the success rate and number of recommendations can be kept high even if the number of wrong recommendations increases, while increasing the number of recommendations.
We used the mean average precision~(MAP)~\cite{moonen2016practical,pugh2018case,rolfsnes2016generalizing,moonen2018effects} and the number of wins and losses~\cite{pugh2018case}, which are also used in other change recommendation studies, to compare the performance of the branch handling strategies.
The detailed definitions of evaluation metrics will be explained later.

In addition, we compare the results for the repository sets used by Kovalenko \etal and those used by Miura \etal, to investigate the influence of repository selection in the experiments.

\subsubsection{Study Design}

We compare the recommendation results between the \Full and \FirstParentNM strategies~(\ImpNM) and those between the \Full and \FirstParentM strategies~(\ImpM).
An overview of the conducted experiment is shown in Figure~\ref{f:expFlow}.
For each repository and each commit to be experimented upon, we created a test set by selecting one file in the changeset as the oracle and using the remaining files as the query.
With the prepared test set, a change recommendation was performed for each branch handling strategy.
Finally, the results based on the branch handling strategies are compared and summarized.

We selected the commits to be used that satisfy the following conditions, which are the same as those used by Kovalenko \etal
\begin{itemize}
  \item At most, 10 files are changed in the commit.
  \item Different branch handling strategies produce different changesets to ensure that the comparison is meaningful~\cite{kovalenko2018mining}.
  \item At least five changesets are collected in either of the branch handling strategies to be able to produce meaningful rules~\cite{kovalenko2018mining}. 
  \item At least one association rule is generated in either of the branch handling strategies~\cite{kovalenko2018mining}.
\end{itemize}

Note that \ImpNM and \ImpM have the following three minor differences in their processing, in addition to the branch handling strategy used.
\begin{itemize}
  \item \textbf{Different restrictions on commit extraction.}
  The commit extraction process is eligible for generating association rules.
  In \ImpNM, up to 100 commits, containing query changes as well as at most 10 changed files in descending order of the authored date~(newer to older), are targeted.
  In \ImpM, for each change targeted in the query, it extracts at most 100 commits using the \texttt{git-log} command with respect to the structure of the commit graph, and commits comprising more than 10 changed files are filtered out.
  \ImpNM fetches commits until it reaches 100, or there are no additional commits with changes matching the query; meanwhile, \ImpM fetches up to 100 commits from the sliced history of each file in the query, and extracts commits comprising at most 10 file changes.
  We made this change for efficiency in \ImpM, as we needed to process large-scale test sets.
  \item \textbf{Different target repository sets}.
  The implementation of Kovalenko \etal excludes repositories containing more than 10,000 commits.
  In \ImpNM, we also limited the repositories that the original implementation of Kovalenko \etal could process, because we wanted to compare the results of our \ImpNM and Kovalenko \etal's implementation to identify the differences.
  \item \textbf{Different adjustment for the number of association rules.}
  The implementation of Kovalenko \etal (\ImpNM) included adjusting the number of commits to be used to generate the association rules and the number of association rules to be extracted to ensure that the results of both branch handling strategies produce the same number of recommendations for fairness.
  This is an unnecessary process in the actual change recommendation process.
  To align with the actual change recommendation process, \ImpM did not follow such an adjustment.
  In contrast, \ImpNM continued to perform the same adjustment because this study focused more on replicating the experiments of Kovalenko \etal
\end{itemize}

The obtained recommendation results are thus compared with several evaluation metrics.
\begin{itemize}
  \item \textbf{Success rate}~\cite{kovalenko2018mining} classifies the recommendation results into \emph{success}, \emph{failure}, and \emph{no prediction}, and summarizes their ratios for each branch handling strategy.
  Note that we have changed the classification criteria differently than Kovalenko \etal
  They regarded a recommendation as a \emph{failure} only when the files included in the query were recommended, whereas we regarded such cases as \emph{no prediction}.
  In addition to the success rate, the \emph{failure rate} and \emph{no prediction rate} were computed.
  \item \textbf{MAP}~\cite{pugh2018case} is used in two ways.
  \MapAll strictly treats the average precision~(AP) of the no-recommendation result as 0, as it assumes that the recommender should always produce at least one recommendation file.
  In contrast, \MapApp excludes such no-recommendation results, as it allows the recommendation system to skip its recommendation results.
  \item \textbf{Number of wins/losses}~\cite{pugh2018case} is calculated by determining the win or loss for each test set.
  A strategy wins if it produces a higher AP than the other strategy, or if it does not make a false recommendation when the APs of both strategies are 0.
\end{itemize}

\begin{table}[t]\centering
  \caption{Summary of differences between two studies}\label{t:rq1}
  \includegraphics[width=8.5cm]{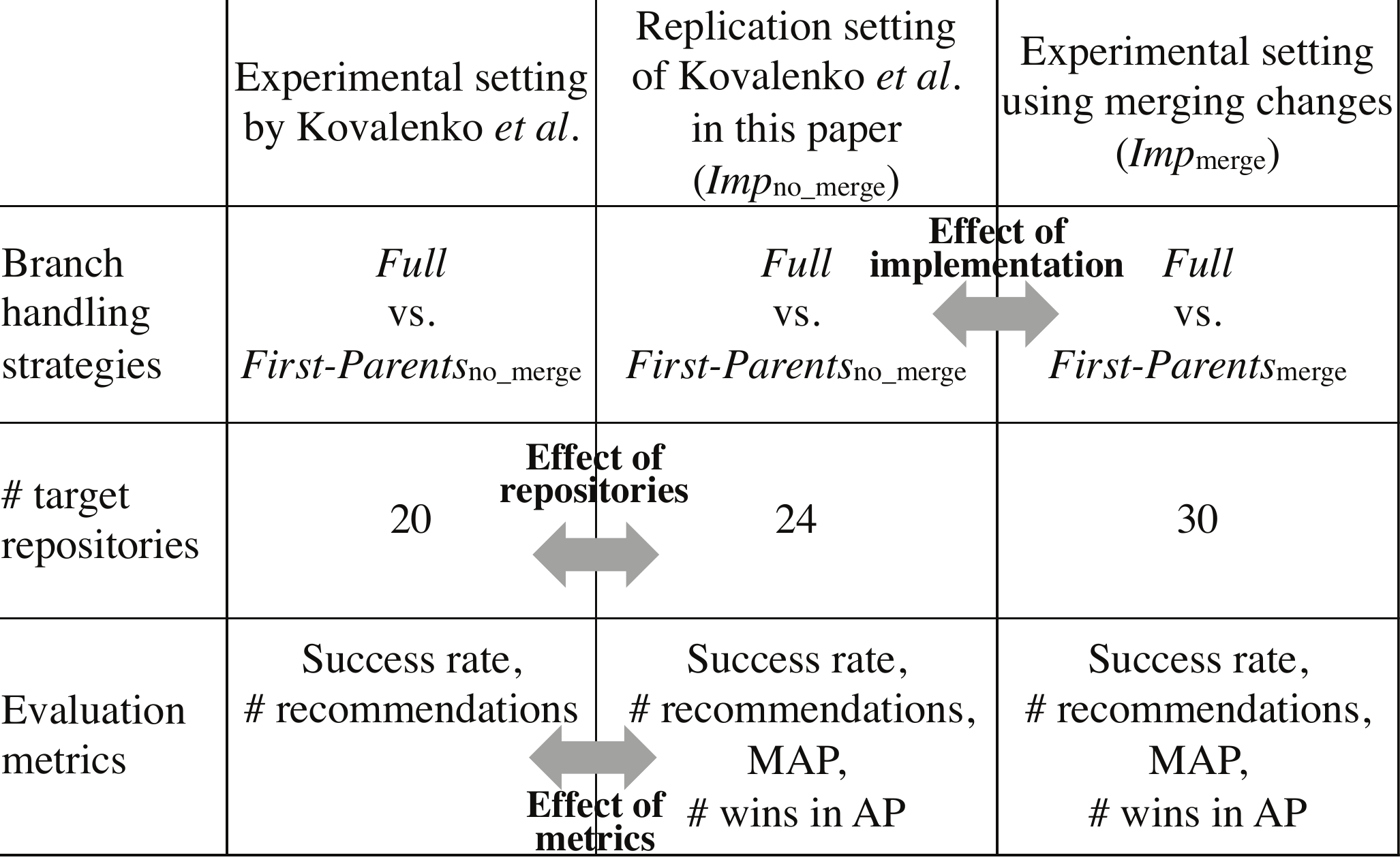}
\end{table}
Table~\ref{t:rq1} summarizes the differences between the study conducted by Kovalenko \etal and the results from answering \RQ{1}.

\begin{table*}[t]\centering
  \caption{Success, failure, and no prediction rates per repository sets}\label{t:ResultGroup}
  \begin{tabular}{cllrrrrrr} \hline
    Implementation & Repositories & Strategy &  Events & Recommendations & Rules     & Success & Failure & No prediction \\
                   &              &          &  count  & (average)       & (average) & rate    & rate    & rate \\ \hline
    \multirow{6}{*}{\ImpNM} & \multirow{2}{*}{\ApacheK}  & \Full          &  25,588 &\W{1.005}& 7.763 &\W{0.179}& 0.438 & 0.383 \\
                            &                            & \FirstParentNM &  25,558 &   0.954 & 7.763 &   0.167 & 0.428 & 0.405 \\ \cline{2-9}
                            & \multirow{2}{*}{\EclipseK} & \Full          &   4,405 &   0.775 & 6.862 &\W{0.156}& 0.372 & 0.472 \\
                            &                            & \FirstParentNM &   4,405 &\W{0.788}& 6.862 &   0.148 & 0.383 & 0.470 \\ \cline{2-9}
                            & \multirow{2}{*}{\ApacheM}  & \Full          &   7,063 &\W{0.726}& 6.491 &\W{0.104}& 0.395 & 0.501 \\
                            &                            & \FirstParentNM &   7,063 &   0.661 & 6.491 &   0.099 & 0.359 & 0.542 \\ \hline
    \multirow{6}{*}{\ImpM}  & \multirow{2}{*}{\ApacheK}  & \Full          &  33,728 &   0.787 & 6.808 &   0.143 & 0.350 & 0.507 \\
                            &                            & \FirstParentM  &  33,728 &\W{0.823}& 7.172 &\W{0.147}& 0.365 & 0.488 \\ \cline{2-9}
                            & \multirow{2}{*}{\EclipseK} & \Full          &   5,994 &\W{0.740}& 6.517 &\W{0.150}& 0.350 & 0.499 \\
                            &                            & \FirstParentM  &   5,994 &   0.739 & 6.792 &   0.148 & 0.349 & 0.503 \\ \cline{2-9}
                            & \multirow{2}{*}{\ApacheM}  & \Full          & 105,807 &   0.739 & 5.391 &\W{0.165}& 0.319 & 0.515 \\
                            &                            & \FirstParentM  & 105,807 &\W{0.745}& 5.516 &   0.163 & 0.325 & 0.512 \\ \hline
  \end{tabular}
\end{table*}

\subsubsection{Results} ~

\Heading{Effect of repository selection}
We investigated the effect of target repository selection on the performance of the change recommendation.
The results for each repository set are shown in Table~\ref{t:ResultGroup} using the same format as the table in Kovalenko \etal’s study, whose columns indicate the implementation used, strategy used, number of test sets to be used (event count), average numbers of recommendations and rules, and evaluation metric values.
In \ImpNM, both the number of recommendations and success rate were higher than those when using the \Full strategy, except for the number of recommendations in \EclipseK.
The results indicate that using different repository sets (\ApacheM), rather than the ones used by Kovalenko \etal (\ApacheK and \EclipseK), did not change the trend of the success rate. Hence, we conclude that \emph{the difference between the two studies was not due to the repository selection}.

\begin{table*}[t]\centering
  \caption{Study results}\label{t:allResult}
  \begin{tabular}{llrrrrr} \hline
    Implementation          & Strategy       & Success rate & \MapAll & \MapApp & \# wins & \# draws \\ \hline
    \multirow{2}{*}{\ImpNM} & \Full          & 0.162 & 0.144 & 0.247 & 3,214 & \multirow{2}{*}{36,708} \\
                            & \FirstParentNM & 0.152 & 0.135 & 0.241 & 3,407 &                         \\ \hline
    \multirow{2}{*}{\ImpM}  & \Full          & 0.160 & 0.143 & 0.294 & 7,848 & \multirow{2}{*}{130,809} \\
                            & \FirstParentM  & 0.159 & 0.143 & 0.289 & 6,872 & \\ \hline
  \end{tabular}
\end{table*}

\Heading{Effect of evaluation metrics}
To investigate the influence of the evaluation metrics to be used, the results of the success rate, \MapAll, \MapApp, and the number of wins and losses for all repositories are computed, as shown in Table~\ref{t:allResult}.
The rightmost two columns show the number of wins and draws for each branch handling strategy.
For \ImpNM, \Full performed better in most evaluation metrics; however, \FirstParent had more number of wins.
\FirstParent had a higher number of wins for \ImpNM because the number of false recommendations increased along with the number of recommendations in \Full.
We suspect that this is due to the missing changes on the branch in the \FirstParent strategy, which reduced the variation of recommendations and number of recommendations.

\begin{table}[t]\centering
  \caption{Number of repositories of higher performance}\label{t:winnerCnt}
  \begin{tabular}{llrrr} \hline
                            &        & \# \Full & \# draws & \# \FirstParent \\
    Implementation          & Metric & wins     &          & wins \\\hline
    \multirow{3}{*}{\ImpNM} & Success rate & 15 &  7 &  2 \\
                            & \MapAll      &  8 & 16 &  0 \\
                            & \# wins      & 13 &  2 &  9 \\ \hline
    \multirow{3}{*}{\ImpM}  & Success rate & 15 &  3 & 12 \\
                            & \MapAll      &  5 & 20 &  5 \\
                            & \# wins      & 16 &  1 & 13 \\ \hline
  \end{tabular}
\end{table}

Table~\ref{t:winnerCnt} summarizes the number of repositories that performed better than the opponent strategy in terms of the success rate, \MapAll, and the number of wins/losses.
To count the winners in terms of the success rate, we considered it a draw if the rate was equal to that of the other strategy.
To count the winners in terms of \MapApp, we used the Wilcoxon signed-rank test for the distribution of AP values.
The winner was decided when we confirmed statistical significance at the 5\% significance level.
If we could not conclude statistical significance, we considered such cases as draws.
Consequently, for the success rate and \MapAll, the performance of the \Full strategy tended to be higher in \ImpNM.
The \Full strategy produced more winning repositories in terms of the number of wins.

The \ImpNM results also indicate that the \Full strategy produced more false recommendations.
However, the performance of the \Full strategy tended to be higher, and we concluded that the effect of the evaluation metrics was small.

\Heading{Effect of not using changes in branches}
The results of \ImpNM and \ImpM in the tables above were compared to determine the effect of skipping information on the branches.
In all results, the branch handling strategy that resulted in higher performance tended to vary.
In the results of Table~\ref{t:ResultGroup}, the branch handling strategies that exhibited higher performance differed depending on the repository set used.
The \FirstParent strategy produced a higher recommendation count and success rate for \ApacheK, whereas the \Full strategy produced a higher recommendation count and success rate for \EclipseK.
For \ApacheM, the number of recommendations was higher when using the \FirstParent strategy, and the success rate was higher when using the \Full strategy.
The magnitude of the difference was smaller than that for \ImpNM.
In Table~\ref{t:allResult}, the difference between the branch handling strategies became smaller in \MapAll, and the draw rate in terms of the number of wins increased from 84.7\%~(36,708/43,329) to 89.9\%~(130,809/145,529).
Moreover, Table~\ref{t:winnerCnt} shows that \ImpM demonstrated better equilibrium than \ImpNM in terms of the number of repositories that the \Full and \FirstParent strategies produced higher performance.

In summary, the results of Kovalenko \etal~\cite{kovalenko2018mining}, which showed that the change recommendation performed slightly better when commits on a branch were handled, were possibly caused by missing change information on branches.

\Conclusion{
  The change recommendation performance could be affected by the experimental setting in handling merging changes.
  Regardless of the variations of the target repository sets and the evaluation metrics, the \Full strategy tended to demonstrate better performance than \FirstParent strategy in \ImpNM; such variations did not affect the conclusion of Kovalenko \etal
  However, this tendency was balanced when using \ImpM, which includes merging changes when using the \FirstParent strategy.
}

\subsection{\RQ{2}: \RQtwo}

\subsubsection{Motivation}

Branches can be used in multiple ways, and different repositories use them differently.
Although most branches are used to implement features or fix bugs~\cite{zou2019branch}, they are also used to manage releases.
For example, several repositories of interest in this study, such as \texttt{apache/accumulo} or \texttt{apache/cassandra}, prepare branches for each version, and developers work on them in parallel.
Even if a branch is used to implement a feature, the frequency of creating a branch and/or granularity of commits and features might differ depending on the commit policy of the project.

Depending on the characteristics of each branch, there may be different ways of handling the branch that is appropriate for extracting changes in the change recommendations.
For example, if changes are frequently split into smaller fragments and commits are kept fine-grained on a branch in a project, it might be better to treat them as a single merge commit, rather than as multiple smaller commits.
In contrast, in the case of making multiple large changes on a branch and merging them, it might be better to treat each commit on the branch separately to collect changesets for the change recommendation and avoid unnecessary coupling.

\begin{figure}[t]\centering
  \includegraphics[width=\linewidth]{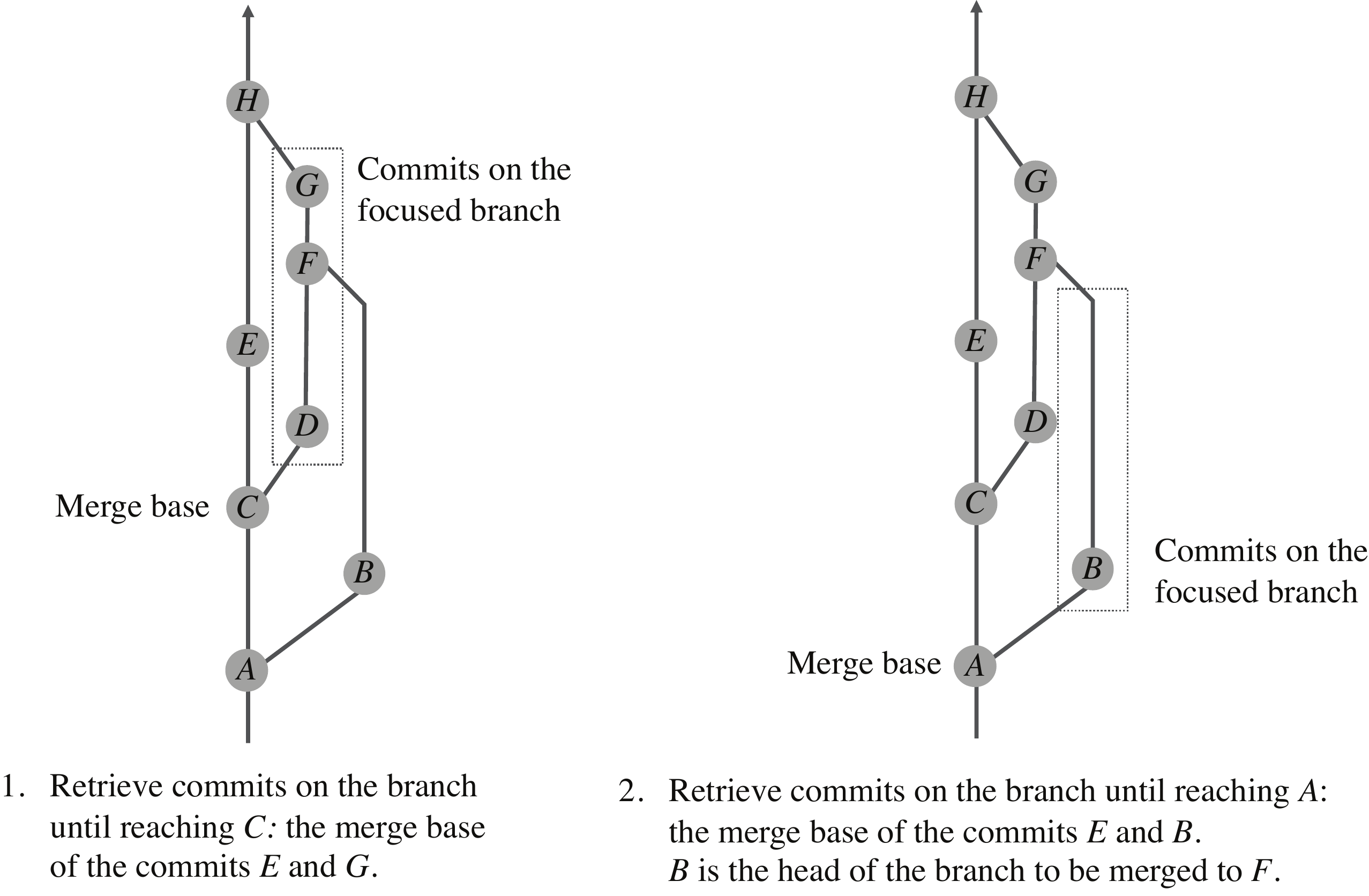}
  \caption{Computation of branch length.}\label{f:BranchLength}
\end{figure}

Therefore, in answering \RQ{2}, we examine the influence of the following two characteristics of a merged branch on the change recommendation performance.
\begin{itemize}
  \item
  \textbf{Branch length} is the number of commits in the path of a branch.
  It counts the commits between the \emph{merge commit} and the \emph{merge base}, \ie, the common parent of the merged branch and the parent branch.
  An illustrative example is shown in Figure~\ref{f:BranchLength}, which shows the computation of the length of the branch merged at commit $H$.
  First, the merge base of this branch, $C$, was computed as the common parent of the head commit of the sinking branch of merge $E$ and source branch of the merge $G$.
  The commits on the target branch until the merge base is read, \ie, $G$, $F$, and $D$, are retrieved.
  When finding another merge commit during this retrieval, we apply this process recursively.
  Because commit $F$ is a merge commit on the acquired branch, we compute the commits on the branch to be merged at $F$ up to merge base $A$. Commit $B$ is newly retrieved.
  Finally, all meaningful commits to be merged at commit $H$, excluding merge commits, are $G$, $D$, and $B$, and its branch length is computed as 3.
  \item
  \textbf{Merge commit size} indicates the size of the changeset of a merge commit, \ie, the number of files that have been changed by the merge commit.
\end{itemize}

We divided this RQ into the following two sub questions:
\begin{itemize}
  \item \RQ{2.1}: \RQtwoone
  \item \RQ{2.2}: \RQtwotwo
\end{itemize}
In \RQ{2.1}, we examine, for each branch characteristic, the trend in change recommendation results for each characteristic of the partial change history used to generate association rules caused by a branch.
In contrast, \RQ{2.2} examines the accuracy of the co-change information of commits on the branch, and the merge commits indicate future changes for each branch characteristic.
The analysis for \RQ{2.1} focuses on the branch that caused the difference in commits used for change recommendation, whereas the analysis for \RQ{2.2} directly compares the co-change information of commits on the branch and merge commits.
Therefore, the analysis for \RQ{2.2} was unaffected by the implementation of the change recommendation method.

\subsubsection{Study Design of \RQ{2.1}}

We investigated the recommendation results for each branch characteristic used in association rule generation in the change recommendation.
The commits used to generate the association rules were replaced by switching the branch handling strategy used.
From the replaced commits, we could identify the merge commit that caused the replacement and the commits on the branch that were additionally captured by the replacement.
We also analyzed the trend of the recommendation results for those characteristics.
Because there may be more than one merge commit that caused the replacement, we compared two typical cases to determine if a similar tendency could be obtained: 1) the results of a single commit and 2) those of six or more commits. 
Boundary six was used for the third quartile of the number of merge commits that caused the swap of commits in each test set to be analyzed.

To compare the recommendation results obtained from \Full and \FirstParentM strategies, we used the number of wins/losses~\cite{pugh2018case}.
Using the values of the branch characteristics involved, we divided the recommendation results such that the number of samples was as even as possible, and confirmed the branch handling strategy that tends to win.
In the case of six or more merge commits, we split the results by the median value to mitigate the effect of outlier instances related to several merge commits.

Note that we limited the number of commits to be investigated to highlight the changes in the commits used between branch handling strategies.
Specifically, we targeted commits obtained by limiting the median (53) number of retrieved commits to generate association rules via the \FirstParent strategy, resulting in 71,832 commits.

\begin{figure*}[t]\centering
  \includegraphics[scale=0.7]{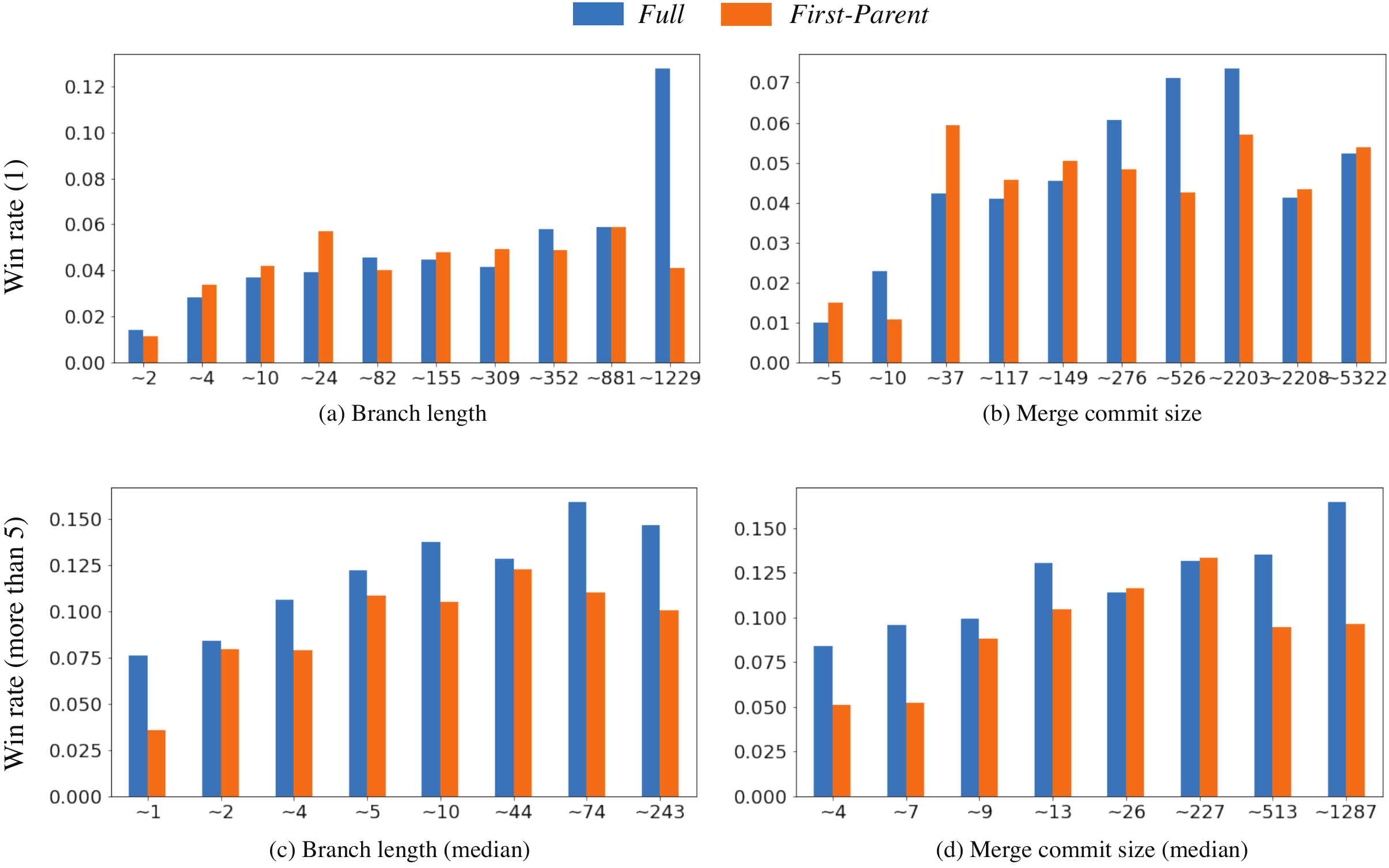}
  \caption{Winning rate of branch handling strategies against branch characteristics.}\label{f:WinnerRate}
\end{figure*}

\subsubsection{Results of \RQ{2.1}} ~

\Heading{Branch length}
The ratio of the winning branch handling strategy to the branch length and the size of the merge commit is shown in Figure~\ref{f:WinnerRate}.
Note that tie cases are excluded in this figure.
In the case of a single merge commit, the effect on the recommendation result increases with the branch length, from 10 to 24.
When the branches become longer than 881, the influence increases.
The difference in the win rate between the branch handling strategies, which was less than 2\% up to a branch length of 881, becomes 8\% higher for the \Full strategy.
If six or more merge commits are handled, the win rate increases until the branch length reaches 5.

Therefore, we conclude that, when the branch is longer, the influence of selecting the branch handling strategy is greater.
If there are extremely long branches involved, they might be better treated separately in the commits on the branch, \ie, using the \Full strategy.

\Heading{Merge commit size}
In the case of a single merge commit, as shown in Figure~\ref{f:WinnerRate}, until the size of the merge commit exceeds 10, there is no winner in approximately 97\% of the cases, and the effect of changing the branch handling strategy is small.
However, when the size exceeds 10, the win rate increases rapidly from 3\% to 10\%.
We believe that this is because the change recommendation method used in this experiment generates an association rule that excludes commits with more than 11 changes.
In the case of six or more merge commits, the rates of winning and losing, rather than draws, tended to increase for winning commits.

As the size of the merge commits increases, the effect of the branch increases.
Additionally, in the case of a single branch, we observe a sudden increase in the win rate when the size of the merge commit exceeds 10.
This is considered to be an effect of the change recommendation method mentioned above.

\subsubsection{Study Design of \RQ{2.2}}

In answering \RQ{2.2}, we compared the co-changed files extracted from a merge commit as a single changeset and those extracted from multiple changesets from the commits on the related branch.
In the analysis, for every single file that is modified by a merge commit or a commit on a branch that is fixed, its co-changed files ($F_\mathit{changed}$) are compared.
The files that are co-changed with the target file in the reachable future 100 commits from the merge commit are regarded as the oracle ($F_\mathit{oracle}$).
By comparing the two file sets, \Precision is calculated as
\[
  \mathit{Precision} = \frac{|F_\mathit{changed} \cap F_\mathit{oracle}|}{|F_\mathit{changed}|}.
\]
To mitigate the effect of the merge commits containing several files, which should be larger than that of smaller merge commits, we obtained the average for each commit, and determined the distribution and trend for each characteristic of the branch.

We targeted all merge commits on the default branch, \ie, the merge commits that can be reached from the repository's HEAD by following the first parent commit, which can be reachable using both the \FirstParent and \Full strategies.
We excluded merge commits where the associated branch length is 1, and the change in the changed files in the branch is the same as that in the merge commit, because the difference in the co-change extraction strategies does not affect such merge commits.

\begin{figure*}[t]\centering
  \includegraphics[scale=0.7]{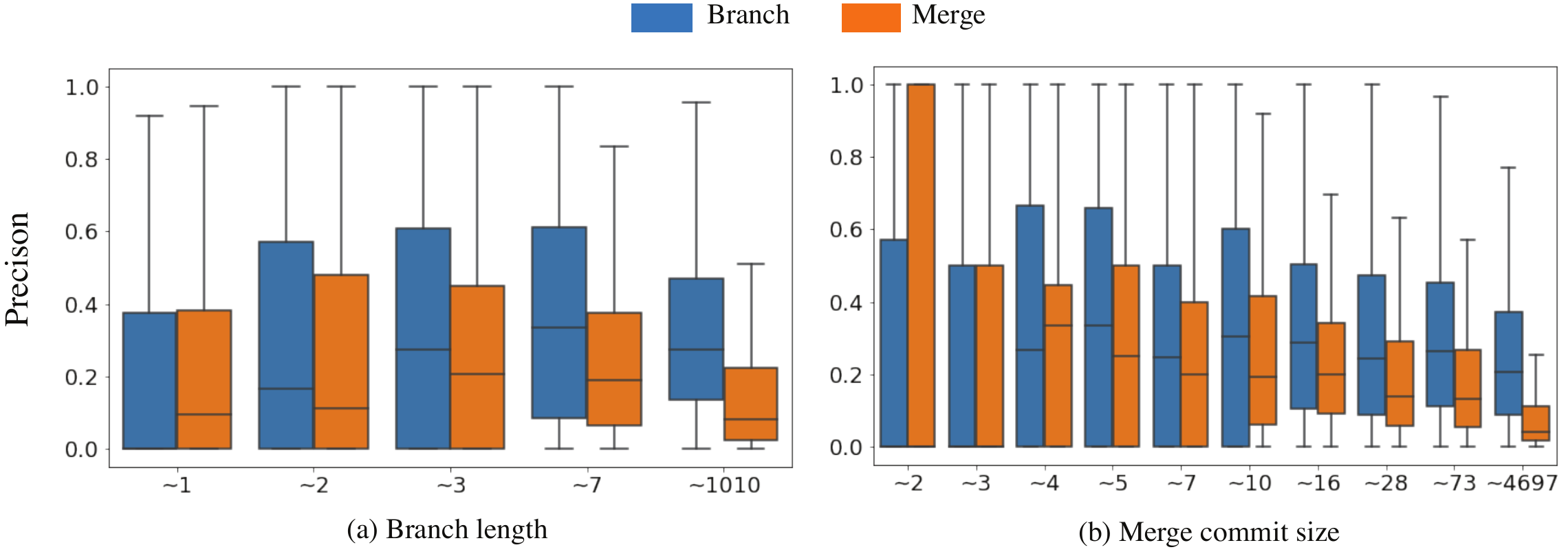}
  \caption{Distribution of \Precision over branch characteristics.}\label{f:coChangeBox}
\end{figure*}

\subsubsection{Results of \RQ{2.2}} ~

\Heading{Branch length}
The distribution of the average values per commit of \Precision for the branch features is shown in Figure~\ref{f:coChangeBox}(a).
Except for branch lengths of 1, the distribution exhibits higher precision when using branches than when using merge commits, and the difference between them increases as the branch length increases.
In cases with a branch length of 1 or less, the \Precision was higher when using merge commits because using branches can only outperform if additional changes, or deletion of unnecessary changes, have been made during the merge.

In summary, except for branch lengths 1 or less, the precision values were higher when the changesets were extracted individually with the commits on a branch. When the branch is longer, the difference is larger.

\Heading{Merge commit size}
As shown in Figure~\ref{f:coChangeBox}(b), when the size of the merge commits is small, the median and third quartiles are higher when using merge commits.
However, as the size increases, the distribution of precision increases when branches are used.
The median and third quartiles are higher when using merge commits, while the merge commit is small. As the merge commit becomes larger, the median becomes higher when using branches.
As merge commits become larger, the extracted co-changes are summarized in a merge increase, and the ratio of not useful changes that will not happen in the future increases.

\subsubsection{Summary}
In both analyses of \RQ{2.1} and \RQ{2.2}, when the branch is longer or the merge commit is larger, the evaluation result is better for treating them separately in commits on a branch.
In the case of longer branches and larger changes, there might be multiple changes belonging to different intentional tasks, and merging them together in a single changeset of a merge commit tends to have a negative effect.
In the case of shorter branches and smaller changes, a single changeset extracted from a merge commit tends to be better for the source of co-changes.
However, these differences were hardly reflected in the recommendation results.
We examined the recommendation results where the maximum length of the involved branches was at most 1, and 98\% of their cases resulted in a tie, although the \FirstParent strategy demonstrated a higher performance in general.
The need for the branch handling strategy to be changed is minute in repositories where branches are used infrequently, and where they are short, because the influence of only a single short branch is negligible for selecting a branch handling strategy.

A sudden increase in the influence of the branch handling strategies occurred when the number of merges exceeded 10 in the recommendation results.
This is because the change recommendation method to be used limited the source commits in generating association rules with at most 10 changes.
When analyzing the co-changes~(\RQ{2.2}), there was no significant differences when the size of the merge commits was approximately 10.
The influence of the branch handling strategies increased because the co-change of the merge commits was not handled anymore.

\Conclusion{
  When longer branches and larger merge commits are involved, the effect of the branch handling strategies becomes more significant, and the performance improves when the co-changes are extracted individually from the commits on a branch.
  In contrast, when shorter branches and smaller merge commits are involved, the performance improves when the co-changes are extracted from a merge commit at once. However, the observed difference was small; the effect was also small.
  The size parameter of excluding commits in the change recommendation method affected the difference in the recommendation results.
}

\section{Discussion and Implications}\label{s:discussion}

\subsection{Discussion}

The results of \RQ{2} showed that, as the branches become longer and the merge commits become larger, handling co-changes with commits on the branch exhibit better performance.
However, if we refer to Figure~\ref{f:WinnerRate}, it is evident that the rate at which the \Full strategy wins does not increase unilaterally; however, the rate at which the \FirstParent strategy wins also increases.
Therefore, we sampled and investigated a small number (40) of merge commits to determine the co-changes added.
Regarding the sampling, we sampled merge commits where the number of co-changes added by the merge is seven or more, to exclude a small number of co-changes that were highly affected by a single co-change and cases where the maximum number of co-changes that a change to three files can have was six or less.
A typical case is the co-change with \texttt{CHANGE.txt} in \texttt{apache/cassandra}, which records the updates.
In this repository, 19 out of 20 merge commits corresponded to this.
The other repositories also contained three co-changes with configuration files.
Additionally, co-changes from newly added changes in the merge commit occurred in 6 of the 40 cases.
For merge commits with additional changes, the probability of future co-changes of the merge was higher than that of co-changes on the branch, which is the opposite of the result for the merge commits without additional changes.
If there are additional changes in a merge commit, it might be better to handle a single changeset extracted from the merge commit.

\subsection{Implications}

In change recommendation, it is suggested that handling commits on a branch separately is better, instead of merging them in a merge commit.
When the branch is longer, it tends to be better at handling commits individually on the branch.
When the branch is shorter, such as those with a length of 1, it is better at handling commits as a merge commit; however, the effect of this treatment is extremely small.
Additionally, because long branches contain changes to many different files, there are many opportunities for a single branch to affect the change recommendation.
Therefore, when deciding whether to handle long branches uniformly as multiple commits on a branch or as a single merge commit, the disadvantage of generating co-changes that will not be made in the future leads to a higher risk than the other strategy, which may lack some co-changes.

However, when using a file as a query that has been modified significantly, when branches are rarely used, or when only shorter branches are used in the target repository, the importance of selecting the branch handling strategies decreases.

\section{Threats to Validity}\label{s:threats}

\Heading{Internal validity}
We studied the change recommendation method based on the ``other files'' algorithm and the implementation by Kovalenko \etal~\cite{kovalenko2018appendix}.
Because the method was designed to limit its recommendations, the effect of selecting the branch handling strategies might be smaller in observation than in reality.

The analysis conducted in this study assumes that changes on a branch are related to a single task in most cases, which is the most typical usage of a branch.
However, this is not always true; the changes in a branch may be related to multiple tasks.
In fact, in the repository of \texttt{apache/cassandra} in \ApacheK, commits related to multiple tasks, where multiple issues are linked, were merged at once.
Therefore, we believe that the results of this study do not negate the results of Miura \etal~\cite{miura2016impact}.

\Heading{External validity}
In our study, only the repositories of \ApacheK and \EclipseK used by Kovalenko \etal~\cite{kovalenko2018mining} and those of \ApacheM used by Miura \etal~\cite{miura2016impact}, were targeted.
The selection of repositories may still be biased; it may have been influenced by their development styles.
For example, in \texttt{apache/cassandra}, where merging was heavily used, the changes corresponding to an issue were combined into a single commit without recording the details of the fine-grained development history.
In such repositories, where the fine-grained development process is performed on branches, using commits on a branch separately, as the source of changesets to be used for the change recommendation, might produce better results.

\section{Related Work}\label{s:related}

Several change recommendation methods that utilize change coupling have been proposed. These include ROSE by Zimmermann \etal~\cite{zimmermann2005mining}, an FP-Tree-based approach by Ying \etal~\cite{ying2004predicting}, CO-CHANGE by Kagdi \etal~\cite{kagdi2006mining}, and TARMAQ by Rolfsnes \etal~\cite{rolfsnes2016generalizing}.
In these methods, changesets obtained from the commits in a repository are used to generate the recommendation rules.
The branch handling strategies studied herein affect the accuracy of these methods.

Moonen \etal studied the various factors that affect the change recommendations.
They investigated the effect of measures, such as confidence and support, which are used to rank association rules. They also studied the effect of limiting the number of file changes in a commit used to create association rules~\cite{moonen2016practical}, that of the history length used, and that of using the most recent history~\cite{moonen2018effects}.
Pugh \etal~\cite{pugh2018case} improved the accuracy of change recommendations based on these results to produce shorter partial histories for generating recommendation rules.

Moreover, Rolfsnes \etal proposed several approaches to improve the accuracy of change recommendation methods, including the aggregation of multiple recommendation rules~\cite{rolfsnes2016improving}, filtration of recommendations via machine learning~\cite{rolfsnes2017predicting}, and use of method-level finer-grained co-change information~\cite{rolfsnes2018aggregating}.
Mondal \etal proposed a combinational approach for analyzing the relationship of program identifiers~\cite{mondal2014improving}.

Zou \etal~\cite{zou2019branch} conducted a study on branch usage in 2,923 repositories, published on GitHub.
Further, Shihab \etal~\cite{shihab2012effect} investigated the influence of branches on software quality for repositories in Microsoft.
Bird \etal~\cite{bird2012assessing} reduced development delays by identifying high-cost branches that require conflict resolution for merging or a long time for validating and deleting branches with high cost, but small contributions.
In contrast to their studies, in this study, we analyzed the effect of each branch characteristic on the change recommendations.

\section{Conclusion}\label{s:conclusion}

In this study, we analyzed the effect of branches on change recommendation results.
In contrast to the study by Kovalenko \etal~\cite{kovalenko2018mining} with a slightly higher performance for the \Full branch handling strategy than the \FirstParentNM strategy, the comparison with the \FirstParentM strategy resulted in a balanced performance among the branch handling strategies.

Additionally, the analyses of branch influence on the change recommendation demonstrated that it is better to treat commits on branches separately in the change recommendation.
Meanwhile, handling all commits on a branch as a single changeset at once in the related merge commit prevented the leakage of co-changes. Performing this treatment on a large branch caused a negative effect of noise from unnecessary co-changes that will not be changed in the future.
We conclude that it is less risky to handle co-changes in the commits on a branch.

Our future work will involve improving the generality of our analysis by experimenting on a wider range of repositories with modern change recommendation methods, such as TARMAQ~\cite{rolfsnes2016generalizing}.

The supplemental materials are publicly available~\cite{appendix}.

\section*{Acknowledgments}
This work was partly supported by JSPS Grants-in-Aid for Scientific Research JP18K11238, JP21K18302, JP21KK0179, and JP21H04877.



\end{document}